\documentclass[twocolumn,showpacs,preprintnumbers,amsmath,amssymb]{revtex4}

\usepackage{graphicx}
\usepackage{dcolumn}
\usepackage{bm}
\usepackage{color}

\begin{document}


\title{Characterizing Topological Order in Superconducting Systems}
\author{M. Cristina Diamantini}
\email{cristina.diamantini@pg.infn.it}

\author{Pasquale Sodano}
\email{pasquale.sodano@pg.infn.it}
\affiliation{%
 Dipartimento di Fisica, University of Perugia, via A. Pascoli, I-06123 Perugia, Italy \\
 and\\
  Sezione INFN, Perugia, Italy
}%


\date{\today}

\begin{abstract}

 Two established frameworks account for
the onset of a gap in a superconducting system: one is  based on
spontaneous symmetry breaking via the Anderson-Higgs-Kibble
mechanism, and the other is based on the recently developed
paradigm of topological order. We show that,  on manifolds with
non trivial topology, both mechanisms yield a degeneracy of the
ground state arising only from the {\it incompressibility} induced
by the presence of a gap.  We compute the topological
entanglement entropy of a topological superconductor and argue
that its measure allows to {\it distinguish} between the two
mechanisms of generating a superconducting gap.

\end{abstract}
\pacs{11.10.z;74.20.Mn;05.30.Pr}

\maketitle

\section{Introduction}
The discovery of the fractional quantum Hall liquids (FQHLs)
\cite{Lau} led to a new understanding of strongly correlated
electron systems and to the idea of a new type of quantum order
\cite{wen}, called topological order. Quantum order in general
describes the zero-temperature properties of  strongly entangled
quantum ground states  not arising from spontaneous symmetry
breaking (SSB). A very surprising property exhibited by FQHLs is
that, due to repulsive interactions and strong correlations
between electrons, they give rise to an incompressible state even
when the first Landau level is only partially filled at some
"magical" filling fraction and all bulk excitations have a finite
gap; no order parameter is available for this state. To explain
these features Wen \cite{wen} introduced the paradigm of
topological order as a special type of quantum order whose
hallmarks are the presence of a gap for all excitations
(incompressibility) and the degeneracy of the ground state on
manifolds with non-trivial topology. Due to incompressibility, at
low energies, the dynamical degrees of freedom of  a FQHLs are
only gapless edge excitations, which may be used to characterize
the various topological orders \cite{wen}. For fractional quantum
Hall systems another important hallmark is the presence of
excitations carrying fractional charge and fractional spin and
statistics \cite{wilczek}.

The long distance behavior of systems exhibiting topological order
is usually described by incompressible fundamental (e.g.
single-layer FQHLs) or non-fundamental (e.g. double-layer FQHLs)
fluids. The simplest example of a $2D$ topological fluid
\cite{wen} is characterized by a ground-state described by a
low-energy effective action given solely by the topological
Chern-Simons (CS) theory \cite{jackiw}:
\begin{equation}S = k/4\pi\ \int d^3x \ A_{\mu} \epsilon^{\mu \nu \alpha} \partial_{\nu} A_{\alpha } \ ,
\label{csa}
\end{equation}
describing a compact $U(1)$ gauge field $A_{\mu}$ whose dual field
strength $F^{\mu} = \epsilon^{\mu \nu \alpha} \partial_{\nu}
A_{\alpha}$ is interpreted as the conserved matter current; the
degeneracy of the ground state on a manifold of genus $g$ is given
by $k^g$ (or $(k_1 k_2)^g$ if $k = k_1/ k_2$ is a rational number).

In \cite{nsm}  a superconductivity mechanism based on a
topologically ordered ground state has been proposed. The dominant
term in the action describing these superconductors is the
topological BF action \cite{birmi}. The BF action, in (D+1)
dimensions, describes the coupling between an antisymmetric tensor
field $B_p$ and the curvature $F = d A_{D-p}$ of a tensor
$A_{D-p}$ , and it is given by:
\begin{equation} S = k/2\pi\ \int d^{(D+1)} \ B_p \wedge d A_{D-p} \ .
 \label{bf}
 \end{equation}
In its application to superconductivity the action given in
eq.(\ref{bf}) is invariant under P-and T- symmetry and is related
to non-chiral fundamental incompressible fluids. A remarkable
realization of this mechanism of superconductivity is provided by
Josephson junction arrays (JJAs) \cite{dst}. In fact, the action
of JJAs  can be exactly mapped onto an Abelian gauge theory with
two gauge fields describing a current of charges and a current of
vortices coupled by a mixed Chern-Simons term \cite{freedman}:
\begin{equation}{\cal L}=-{1\over 4e^2}F_{\mu \nu}F^{\mu \nu } + {\kappa \over 2\pi }
A_{\mu }\epsilon ^{\mu \alpha \nu }\partial_{\alpha }B_{\nu }-{1\over 4g^2}
f_{\mu \nu }f^{\mu \nu } \ .
\label{mcs}
\end{equation}
We refer to this model as a fundamental BF model, since
(\ref{mcs}) is derived from the microscopic model and does not
describe only the low energy degrees of freedom as is the case of
effective theories. The Abelian gauge theory exactly reproduces
the phase diagram of JJAs and the insulator/superconductor quantum
phase transition at T=0 \cite{dst}. The insulating phase turns out
to be dual to the superconducting phase and it is also
topological. Global superconductivity in planar JJA is thus the
simplest example of the mechanism of superconductivity proposed in
\cite{nsm}. JJA  have also been recently  considered by several
other authors \cite{as},  as controllable devices exhibiting
topological order. Of course, planar JJAs with open boundary
conditions do not exhibit ground state degeneracy. To see a ground
state degeneracy one should require periodic boundary conditions
implementing  a torus topology; {\it toroidal} arrays are not
difficult to realize in practice \cite{cirillo}.

Abelian effective CS theories are only one of the two
candidates to describe the long-distance properties of fundamental
topological fluids: the other is given by $W_{1+\infty}$ minimal
models; indeed, Abelian CS theories describe special cases of
quantum incompressible fluids, but not all of them. As evidenced
in \cite{ctz}, the $W_{1+\infty}$ minimal models are not described
by Abelian  CS theories since, while all the Abelian CS theories
have a dynamical $W_{1+\infty}$ symmetry algebra, the contrary is
not true.

 It should be stressed  that $W_{1+\infty}$ models imply
non-Abelian fractional statistics even for fundamental quantum
Hall fluids, although they are not described by non-Abelian CS
models.  Intuitively, a model of a topological fluid based on
the $W_{1+\infty}$ algebra is "more incompressible" than its CS
counterpart since it supports less degrees of freedom and yields a
different entanglement entropy. Non-Abelian CS models have been proposed, indeed, as effective theories for
composite, non-fundamental quantum incompressible fluids, e.g.
double-layer quantum Hall systems; as such they are not best
suited to study the fundamental mechanism at the origin of quantum
incompressibility. As evidenced in \cite{bulga}, non-Abelian CS
models can be viewed as cosets of $W_{1+\infty}$ fundamental
models and this naturally reflects  their composite character.

Topological field theories emerge naturally also in describing
superconductors. It is well known \cite{dewild} that the degrees
of freedom of an
 Abelian $U(1)$ Higgs model spontaneously broken to $Z_N$ ($N=2$ for conventional superconductivity) are charges $n e\  (n= 0 ...... N-1)$ and
 vortices ${k 2 \pi \over Ne} (k = 0.....N-1)$. While Coulomb interactions are screened in the broken phase, Aharonov-Bohm interactions are
 not and the latter are described by {\it effective} BF theories described by the action given in eq.(\ref{bf}).
 In (2+1) dimensions
 eq.(\ref{bf}) is the action of a mixed CS
model eq.(\ref{csa}); here, the gap and the ground state
degeneracy are due to the breaking $U(1) \rightarrow Z_N$.

Mixed CS models have recently attracted a wide interest
\cite{csto} to describe topological fluids  in various contexts.
As already mentioned, Abelian mixed Chern-Simons models emerge
naturally as an effective theory of the low-energy long -distance
properties of superconductors no matter what is the microscopic
mechanism by which the gap is opened. In this paper we shall
evidence  that, for superconducting systems, quantum
incompressibility alone (i.e. the emergence of a gap) is enough to
account for all the features normally associated with topological
order in $2D$, like the emergence of a ground state degeneracy on
spaces with non trivial topology and of excitations with
fractional charge and statistics. The gap can be of different
origin: SSB or topological, but we will show that this has no
influence on the kinematics of the resulting incompressible fluid.

There is a difference in the ground state degeneracy between
topological and standard superconductors \cite{cpc} and this
allows to distinguish between the two microscopic mechanisms by
which the superconducting gap is opened: indeed, for topological
superconductors the ground state degeneracy is entirely determined
by the coefficient $k$ of the BF action while, for standard
superconductors, the ground state degeneracy is due entirely to
the SSB of $U(1) \rightarrow Z_N$. As a result, there is a
remarkable difference in the ground state degeneracy exhibited by
the standard superconductors in (2+1) dimensions analyzed in
\cite{hansson} and the topological superconductors analyzed in
\cite{nsm}: this is indeed relevant since, as we show in this
paper,  by computing the entanglement entropy of the ground state
\cite{kitaev}, it becomes possible to distinguish between
topological and standard superconductors.  Since an
experimental footprint of the entanglement entropy may be observed
in transport measurements in strongly correlated systems
\cite{nasa}, one may envisage that similar effects may be
evidenced in toroidal JJA's.

Our subsequent analysis is based on the observation
that, once there is a gap, no matter what is its origin, the
ground state has the universal properties of a fundamental quantum
incompressible fluid characterized by a dynamical $W_{1+\infty}$
symmetry algebra (or $ W_{1+\infty} \otimes \bar W_{1+\infty}$ if
they are not chiral as in the BF model we are interested in) for
the edges excitations both for topological and standard
superconductors. As already pointed out in \cite{pima}, the
symmetry under quantum area preserving diffeomorphisms arising
from the tquantum incompressibility alone is enough to account for
the emergence of both a ground state degeneracy and fractional
spin and statistics. Thus, looking only at the edges excitations
does not provide useful information to distinguish between
topological and standard superconductors; at variance, counting
the ground state degeneracy provides a way to characterize these
two very different microscopic mechanism of opening a
superconducting gap.

In section II we analyze incompressible fluids on a torus: our aim
here is to show that ground-state degeneracy arises only from
incompressibility. For this purpose we show that, no matter how
the gap is opened, once one has an incompressible fluid on a
torus, the emerging dynamical algebra, namely the Fairlie,
Fletcher and Zachos trigonometric algebra \cite{ffz}, is -by
itself- enough to account for the ground state degeneracy; as a
result, every incompressible fluid has ground state degeneracy on
a torus. We evidence how the $W_{1+\infty}$ algebra emerges as a
dynamical quantum symmetry group of Abelian CS theories and how
the $W_{1+\infty} \otimes \bar W_{1+\infty}$ arises for the BF
model. We then argue that, for pure CS theories, it is possible to
relate directly the generators of the large gauge transformations
(determining the ground state degeneracy of a CS theory
\cite{cage}) with those of the area preserving diffeomorphisms.
Finally, we show that the ground state degeneracy is affected by
the how the gap originates.
 In Section III we analyze incompressible fluids on manifolds with
boundaries: there, we show that the Abelian mixed CS theory
possess the full $W_{1+\infty}$ dynamical symmetry algebra.
Moreover, using the holographic partition functions, we show
that,despite the fact that once the gap is opened both
superconductors share the properties of an incompressible fluid, a
measure of the topological entanglement entropy {\it may
distinguish} between the two.
 Section $IV$
is devoted to a few concluding remarks.

\section{Incompressible fluids on a torus}
The configurations of $2D$ incompressible fluids are spanned by
area-preserving diffeomorphisms. As a result, the dynamical
symmetry algebra of quantum incompressible fluids is given by
$W_{1+\infty}$, which is the quantization of area-preserving
diffeomorphisms. In this section we show that, for a torus
topology and for both chiral and non-chiral fluids, the dynamical algebra alone is
responsible for the ground-state degeneracy, independently of the
origin of the gap and that, using the algebra $W_{1+\infty}$
\cite{ctz},  allows  to determine the bulk and edge excitations of
a quantum incompressible fluid.

We start by noticing that incompressibility implies that one may regard
the original coordinate space as the phase space, i.e. $(p,q \equiv
x_1, x_2)$ since, from Liouville' s theorem,  one knows that canonical
transformations should preserve the volume in phase space. On this space, canonical transformations
are defined
as $\delta q = \{q, F(p,q)\} =
\partial F /
\partial p$, $ \delta p = \{q, F(p,q)\} = - \partial F / \partial
q$, where $(x_1, x_2)$ are the space coordinates and $ \{, \} $ is
the Poisson bracket.
On a plane, a basis of generating functions may be given in terms of the
complex coordinates $z = x_1 + i x_2, \ \bar z = x_1 - i x_2$, as
\begin{equation} F^{cl} _{n,m}  =  z^n \bar z^m \  ,  \label{wcp}
\end{equation}
which satisfy the classical $w_\infty$ algebra of area-preserving
diffeomorphisms. On a square torus ( i.e. a torus with equal sides
$2\pi$), instead, the generators are given by a set of complete harmonics
\begin{equation} F^{cl} _{n,m}  = { \rm exp }i \left\{ x_1 n  + x_2 m \right\} \  ,  \label{wct}
\end{equation}
with $n$, $m$ integers. Thus, at the classical level, one gets the
algebra $\{  F^{cl} _{\bar n}, F^{cl} _{\bar m } \} =  {\bar n}
\times {\bar m} F^{cl} _{\bar n +\bar m} $.

To quantize this algebra one may consider a  generic torus of dimensions $ L_x$, $
L_y$ and ( with a pertinent normalization) assume the following
commutation relation
\begin{equation}\left[ {x\over L_x}, {y\over L_y}\right] = {i \over 2 \pi k} \  ,
\label{crt}
\end{equation}
where $k$ is a dimensionless constant giving the ratio between the
volume  of the torus  and the volume  of the unit cell in phase
space. In the following we assume $k = k_1/k_2$, with $k_1$ and $k_2$
integers. If one regards ${x\over L_x}$ as a coordinate and $
{y\over  2 \pi k L_y}$ as a momentum in the phase space, the generators may
be written as
\begin{equation}K_{m,n} =  { \rm exp } 2 \pi i \left\{  {x\over L_x} n  +  {y\over L_y} m \right\} \  ,\label{gt}
\end{equation}
with commutation relations given by
\begin{equation}\left[ K_{m,n} ,K_{p,q} \right] =  2i{\rm sin} \left[  {2 \pi \over k} \left( m q - n p \right) \right]  K_{m+p, n+q} \  ,\label{cgt}
\end{equation}
which is the $W_{1+\infty}$ algebra on the torus, also called the
Fairlie Fletcher Zachos trigonometric algebra \cite{ffz}. States of
the quantum incompressible fluid must then fall into
representations of this algebra.

To study the representations of the  $W_{1+\infty}$ algebra one
observes that
\begin{equation} K_{m,n} K_{p,q} =  {\rm exp} \left[  {2 \pi i \over k} \left( m q - n p \right) \right]    K_{p,q} K_{m,n} \  ;\label{cogt}
\end{equation}
thus, the algebra admits a non-trivial two-cocycle and, if $k =
 k_1/k_2$, the generator  $ K_{k_1 m,k_1 n}$ is a Casimir operator for the
algebra . When $k$ is rational, if one applies the generator
$K_{m,n}$ to a physical state $\psi ( x)$, one gets
\begin{equation}K_{m,n} \psi (x) =  {\rm exp} \left[  {i m x\over k  L_x} \right] {\rm exp} \left[  {m n  \pi i \over k} \right] \psi (x + 2 \pi n)  \  ,
\label{gss}
\end{equation}
so that
\begin{eqnarray}K_{k_1 m,k_1 n} \psi (x) &=&  {\rm exp} \left[  {i m  k_2 x\over   L_x} \right]  \times \nonumber \\
&\times& {\rm exp} \left[  i m n  k_1k_2  \pi\right] \psi (x + 2 \pi k_1  n)  \nonumber \\
&=& {\rm exp}  i \phi_{ k_1k_2}  \psi (x) \  ,
\label{css}
\end{eqnarray}
where $\phi_{ k_1 k_2} =  2 \pi  k_1 k_2( m \theta_1 + n \theta_2)$. In
fact, the action of a Casimir operator on the physical state can
change it only by a phase, which is fully specified by the two
angles $\theta_1$ and $\theta_2$. As a result, one may write the
wave function as
\begin{equation}\psi (x) = \sum_p  {\rm exp} \left[  {i p  x\over   L_x} \right]  {\rm exp} \left[  {i \theta_1  k_2 x\over   L_x} \right]  c(p) \
,
\label{peri}
\end{equation}
where $c(p)$ satisfies the quasi-periodicity condition $ c(p - m k_2
) = {\rm exp} \left[  i \theta_2 m   k_1 k_2 \right]  c(p)$, as it can
be seen by applying the Casimir to $\psi (x)$. Thus, for each $k_1$,
one has $k_2$ independent states implying a degeneracy given by $k_1
k_2$ on the torus while, on a surface of genus $g$, it is given by
$( k_1 k_2 )^g$.
We have thus shown that incompressibility alone leads to degeneracy of the ground state
We point out that this degeneracy is only a kinematic
consequence of quantum incompressibility and has nothing to do
with the dynamical origin of the gap from which
incompressibility arises.

We will now show how   $W_{1+\infty}$ and  $W_{1+\infty} \otimes \bar W_{1+\infty}$ algebras arise generically as
the dynamical symmetry group of Abelian CS theories and BF theories respectively. For the former
we focus on the global time-independent symmetries of the
CS action since they are the symmetries of the configuration
space. Quantum states fall then into representations of this
symmetry group, the action being represented by time-independent
charges.

The largest time-independent symmetry group of a CS gauge theory is
given by global time-independent gauge transformations and spatial
diffeomorphisms.  Since the CS action is topological, for compact
surfaces, only gauge transformations survive. This can be easily
seen if one recalls that gauge fields behave as a phase space and
satisfy commutation relations given by $\left[A_i(x) , A_j(y)
\right] = ( i 2\pi / k) \epsilon_{ij} \delta(x-y)$, where $ i=1,2$
label the space coordinate of the manifold. For CS models defined
on manifolds with boundaries the relationship with a
$W_{1+{\infty}}$ algebra was first established for a torus in
\cite{ian}, using the equivalence between a CS model and the
Landau problem. Here, we follow a different approach and show
that, for  pure CS model, it is possible to relate directly the
generators of the large gauge transformations with those of the
area preserving diffeomorphisms.

We start by decomposing the spatial component of the gauge field
into global and local degrees of freedom. In fact, the
contributions of these two parts decouple in the action. Following
\cite{cage,gord} we decompose the gauge field $A$ into exact,
co-exact and harmonic parts as
\begin{equation}A = d \phi + \delta \chi + h \ ;  \  h = 2 \pi \sum_{i = 1}^g \left( u_i \alpha_i +  v_i \beta_i \right) \
.
\label{dgfh}
\end{equation}
We restrict ourselves to $g=1$ with $\alpha$ and $\beta$ canonical
harmonic 1-forms of the surface (dual to a canonical homology
basis). The commutation relations between the gauge fields imply
that
\begin{eqnarray} \left[ \delta \chi(x), \phi(y) \right] &=& {2 \pi \over i k} \delta^2 (x-y) \nonumber \\
 \left[ v_i, u_j) \right] &=& {1 \over 2 \pi i k} \delta_{ij} \
 \label{crcs}
 \end{eqnarray}
 with $k$ the CS coefficient. Eq.(\ref{crcs}) tells then that $\chi$ and $v$ may be regarded as momenta. Finally, requiring that the Gauss law
 annihilates physical states, one gets a quantum theory which is gauge invariant under
small gauge transformation. Thus, the configuration space is
reduced to a $g$-dimensional (1-dimensional if, as in our
derivation, $g=1$) space with coordinates $u_i$ ($i=1$ when $g=1$).

Large gauge transformations $u \rightarrow u + n; v \rightarrow v+
m$ are generated by
\begin{equation}U_{m,n}  =  { \rm exp } \left\{ { \partial \over \partial u} n  +  2 i \pi k m \right\}  \
\label{gct}
\end{equation}
and satisfy the algebra
\begin{equation}
\left[U_{m,n} , U_{p,q}\right] =  2i  { \rm sin } \left\{    \pi k (np - mq ) \right\}  U_{m+p,n+q} \ ,
\label{compl}
\end{equation}
which is nothing else than (\ref{cgt})  with the identification:
${x\over   L_x} = 2 \pi k u$ and ${y\over   L_y} = - { i  \over 2 \pi k} { \partial
\over \partial u}$. The generators of large gauge transformations
can be identified then with the generators of $W_{1+\infty}$. The
CS coefficient $k$ measures the ratio between the total flux
piercing the torus and the unit flux; when there is
commensurability, namely when $k = k_1/k_2$, the degeneracy of the
ground state is given by $ k_1/k_2$.

Let us now concentrate on the BF model and show how  a $W_{1+\infty} \otimes  \bar W_{1+\infty}$ algebra arises  as
the dynamical symmetry group of Abelian mixed CS theories describing topological superconductors and described by the action:
\begin{equation}
S_{BF} = {k \over 2 \pi} \int_{M_{2+1}} A_1 \wedge d B_1\ .
\label{sbf}
\end{equation}

On the torus and let us decompose the forms $A$ and $B$ in the exact, co-exact  and harmonic part:
$$A = d \phi_A - *d \chi_A + u \alpha + v \beta \ \ , \   B = d \phi_B - *d \chi_B + w \alpha + r \beta \ ,$$
where $\alpha$ and $\beta$ form a basis of harmonic one-forms, dual to a canonical homology basis.
The action eq.(\ref{sbf}) become then
\begin{equation}
2 \pi k \int_R r \dot u + v \dot w \ ,
\label{afh}
\end{equation}
with commutation relations
\begin{equation}
\left[ r, u \right]  =   \left[ v, w \right] = {1 \over 2 i \pi k} \ .
\label{ssc}
\end{equation}
The generators of large gauge transformations can thus be written as:
\begin{equation}
U(n,m,t,l) = \exp \left[ n {\partial \over \partial w} + m {\partial \over \partial u} - 2 \pi k (t w + l u) \right]  \ ,
\label{glg}
\end{equation}
and they satisfies the algebra:
\begin{eqnarray}
&&U(n_1,m_1,t_1,l_1) U(n_2,m_2,t_2,l_2) = \nonumber \\
&&= U(n_1+n_2,m_1+m_2,t_1+t_2,l_1+l_2) \times \nonumber  \\
&&\times \exp \left[ i \pi k (t_1 n_2 + l_1 m_2 - n_1  l_2 -m_1 t_2) \right] \ .
\label{aglg}
\end{eqnarray}
We note that for $k$ an even integer the cocycle is always trivial.

This case corresponds to having two incompressible fluids with coordinates $x_i$ and $ y_i$ ($i = A, B$) corresponding to the two gauge fields $A$ and $B$ and commutation relations
\begin{equation}\left[ {x_{A (B)}\over L_x}, {y_{B (A)}\over L_y}\right] = {i \over 2 \pi k} \  ,
\label{crt}
\end{equation}
with generators
\begin{equation}K_{\bar m,\bar n} =  { \rm exp } 2 \pi i \left\{  {x_i\over L_x} n_i  +  {y_i\over L_y} m_i \right\} \  .\label{ngt}
\end{equation}
With the identification:
${x_A\over   L_x} = 2 \pi k u$ , ${y_A\over   L_y} = - i  { \partial
\over \partial u}$ and  ${x_B\over   L_x} = 2 \pi k w$ and ${y_B\over   L_y} = - i  { \partial
\over \partial w}$, the generators of large gauge transformations can be identified then with the generators of $W_{1+\infty} \otimes  \bar W_{1+\infty}$ algebra.

The generators (\ref{ngt}) can be always be rewritten as:
\begin{equation}K_{\bar m,\bar n} =  K_{n_+, m_+} K_{n_-, m_-} \  ,
\label{diag}
\end{equation}
with
\begin{eqnarray}&& K_{n_+, m_+} =  { \rm exp } 2 \pi i \left\{n_+\left( {x_A\over L_x} +  {x_B\over L_x} \right) + m_+\left( {y_A\over L_y} +  {y_B\over L_y} \right) \right\} \nonumber \\
&&K_{n_-, m_-} =  { \rm exp } 2 \pi i \left\{n_-\left( {x_A\over L_x} -  {x_B\over L_x} \right) + m_-\left( {y_A\over L_y} -  {y_B\over L_y} \right) \right\}  \ .
\label{rcd}
\end{eqnarray}
>From eq.(\ref{diag}) we clearly see that $ K_{n_+, m_+}$ and $K_{n_-, m_-} $ are two sets of commuting operators corresponding to the combination $ {x(y)_A\over L_{x(y)}} +  {x(y)_B\over L_{x(y)}}$ and ${x(y)_A\over L_{x(y)}} -  {x(y)_B\over L_{x(y)}}$ . In this case we thus have a $W_{1+\infty} \otimes  \bar W_{1+\infty}$ algebra thereby giving a degeneracy on a torus $(k_1 k_2)\times (k_1 k_2)$.
This corresponds exactly to the degeneracy we expected for a mixed CS theory when both gauge fields are compact.
In \cite{cpc} we have however shown that the superconducting phase of the mixed CS theory has a ground state degeneracy that is only $(k_1 k_2)$ because due to the confinement of magnetic vortices one of the two gauge fields behaves as non-compact.
>From the point of view of the $W_{1+\infty} \otimes  \bar W_{1+\infty}$ algebra, this corresponds to the fact that only one of the two chirality survives.

Superconducting systems allow to realize that, even if a ground state degeneracy
should be always present due to the emergence of a gap (incompressibility), there is still a remarkable difference
between models in which the gap arises as a result of SSB and models where it originates from topology;
in fact, for the latter situation, the ground state degeneracy is characteristic
of the theory independently of which phase is realized while, in the former situation, the ground state degeneracy
appears only after SSB.

There is also another fundamental difference between these two superconductors.
When the gap arises from SSB, the degeneracy is determined by the breaking $U(1) \rightarrow Z_N$:
in \cite{hansson} the $U(1)$ symmetry is spontaneously broken down to $Z_2$
by Cooper pair formation and the ground state degeneracy is $4$; the residual Aharonov-Bohm interactions
between charges and vortices can be described here by an effective mixed CS model with $k=2$ \cite{hansson}.
The degeneracy $k^2$ corresponds, in the topological BF model \cite{cpc}, to the case in which both topological defects, electric and magnetic,
are in a dense phase. However this phase is not allowed dynamically and topological superconductors corresponds to a phase in which only an electric condensate is present and the ground state degeneracy is $k$ or $k_1 k_2$ if $k = k_1/k_2$.  For instance, since planar JJAs are described by a fundamental
CS theory with two coupled gauge fields \cite{dst}, one finds that the degeneracy on a torus is one for the unfrustrated
model, which is described by a mixed CS model with CS coefficient $k = 1$, and  two for the fully frustrated model,
which has $k = 1/2$ \cite{nsm}.

Since the degeneracy on the torus is equivalent to the number of particle types \cite{freedman}  and the quantum dimension of each type of particle is one for Abelian systems, this difference in the ground state degeneracy between the two superconductors leads to a {\it different topological entanglement entropy}     \cite{kitaev} in the ground state, as we will show with a more formal argument in the next section.
In fact  topological entanglement entropy is defined as  \cite{kitaev}  $-\gamma = {\rm log {\cal D}}$ where $ {\cal D}  = \sqrt{\sum_a d_a^2}$ is the total quantum dimension and $d_a$ is the quantum dimension of a particle of charge $a$, and for a system with $k$ or $k^2$ type of particles each of quantum dimension $d = 1$,  we will have $\gamma = \log \sqrt k$ and  $\gamma = \log k$ respectively. For topological superconductors the entanglement entropy is thus half of the one of superconductors originating from SSB. A measurement of the topological entanglement entropy can thus distinguish between the two.

\section{Manifolds with boudaries}

On a disk $D$, the classical  generators of area preserving
diffeomorphism algebra, $w_\infty$,  are given by eq.(\ref{wcp}).  At
the quantum level the generators of $w_\infty$, $W_{1+\infty}, \
V_n$, are characterized \cite{kac} by a mode index $n \in Z$ and
a conformal spin $h = i+1$, and satisfy the algebra:
\begin{eqnarray}& \left[ V^i_n, V^j_m \right] &= (jn -im ) V^{i+j-1}_{n+m} + q(i,j,m,n)  V^{i+j-3}_{n+m}  + \nonumber \\
&. \ . \ . \ . +& c \delta^{ij} \delta_{m+n,0} d(i,n)  \  ,
\label{gaq}
\end{eqnarray}
where $q$ and $d$ are pertinent \cite{kac} polynomials and $c$ is
the central charge. The dots stand for a series of terms involving
the operators  $ V^{i+j-1-2k}_{n+m} $ \cite{kac}.

The generators $ V^{0}_{n}$ are associated to the charge  of the
edge excitations while $V^{1}_{n} $ to the angular-momentum modes;
they satisfy the Kac-Moody algebra:
\begin{eqnarray}\left[ V^0_n, V^0_m \right]  &=& n c \delta_{m+n,0}  \  , \nonumber \\
\left[ V^1_n, V^0_m \right] &=& -m  V^0_{n+m} \ , \nonumber \\
\left[ V^1_n, V^1_m \right]  &=& (n  -m ) V^1_{n+m} + {1  \over 12} c n (n^2 -1) \delta_{m+n,0} \ .
\label{akm}
\end{eqnarray}

All $W_{1+\infty}$ unitary, irreducible, highest-weight
representations have been found by Kac \cite{kac} and applied to
incompressible quantum Hall fluids by Cappelli, Trugenberger and
Zemba  \cite{ctz}. These representations exist only for positive
integer central charge $c=m=1,2,\dots$ and, if $c=1$, they are
equivalent to those of the Abelian sub-algebra $\hat U(1)$ of
$W_{1+\infty}$ , corresponding to the edge excitations of a single
Abelian CS theory. If, instead, $c=2,3,\dots$ there are two kinds
of representations, generic and degenerate, depending on the
weight. The generic representations are equivalent
 to the corresponding representations  of the multi-component Abelian algebra $\hat U(1).^{\otimes m}$ having the same
 weight and correspond to the edge excitations of a multiple Abelian CS
theory. The degenerate representations instead are contained in
the corresponding $\hat U(1)^{\otimes m}$ representations, i.e.
the latter being reducible $W_{1+\infty}$ representations.

Let us consider now a CS theory defined on $D \times R$ (where $R$
accounts for time). It has been shown by Witten  \cite{wit} that
this theory may be quantized upon eliminating only the degrees of
freedom associated to the interior of $D$. As a result, gauge
transformations- which, as already noticed, may be regarded as
diffeomnorphisms \cite{bbgs} in a CS model- relate equivalent fields only in the
interior of the disc, while on $\partial D$ they play a role
somewhat similar to global gauge transformations. The residual
states localized on the circular boundary are the CS edge states
and are equivalent to the edge excitations of a droplet of
incompressible fluid. Furthermore, the generators of the gauge
transformation which do not vanish on $\partial D$ generate a
$U(1)$ Kac-Moody algebra isomorphic to the algebra of a chiral
boson moving on $\partial D$ so that, after quantization, the edge
states satisfy a Virasoro algebra with central charge $c=1$. It
has been shown \cite{ctz} that the generators of this Virasoro
algebra commute with the boundary Hamiltonian and correspond to
local coordinates transformations. In the following we show that
the edge excitations generate not only the Virasoro algebra, but
the full algebra $W_{1+\infty}$: namely, there are more operators
which commute with the boundary Hamiltonian but do not generate
local coordinates tranformations \cite{ctz}.
This new result generalize to manifold with boundaries the results obtained for the torus, where the generators of large gauge transformations are in $1 \leftrightarrow 1$ correspondence with the one of the FFZ algebra.

To prove our result let us review some well known results for edges excitations \cite{wen}.
Following \cite{bbgs}, if one considers gauge transformations
satisfying $\Lambda |_{\partial D} = {\rm exp }(iN \theta) N \in
Z$, the commutators between global charges can be written as
$$\left[ Q_N,Q_M\right]  =  k N \delta_{N+M,0}.$$  This provides a CS construction
of the Fourier modes of a massless chiral boson on a circle. The
Virasoro generators may be written, using the Sugawara
 construction, as
\begin{equation}
L_N= {1 \over 2k}  \sum_{L = -\infty}^\infty  : Q_{N-L} Q_L: \quad  ,
\label{defg}
\end{equation}
and satisfy the algebra
\begin{equation}
  \left[L_N, L_M \right]  =  (N-M) L_{N+M} + {c\over 12} (n^3 -n) \delta_{N+M,0} \ ,
\label{adcv}
\end{equation}
with central charge $c =1$. As a result, the edge excitations of a
CS gauge theory are described by the chiral boson theory
\cite{wen}.

To obtain the action for the chiral boson one may start from the
CS action \cite{wen}:
\begin{equation}S_{CS}  = {k \over 4 \pi} \int  A_\mu \epsilon_{\mu \nu \alpha} \partial_\alpha A_\nu \
,
\label{acs}
\end{equation}
which, in the gauge $A_0 = 0 \implies A_i = \partial_i \phi$,
yields
\begin{equation}S_{EE}  = {k \over 4 \pi} \int \partial_t    \phi \partial_x \phi dx dt \ ,
\label{cbsv}
\end{equation}
where $t$ is the time coordinate and $x$ is the coordinate
parallel to the boundary, which is conveniently parametrized as $x
= R \theta$ with $R$ being the radius of $D$. The equation
describes a chiral boson with zero velocity. This is not
surprising since a finite velocity of the edge excitations is a
boundary effect, which may be induced by a large gauge
transformation. In fact, upon fixing the gauge as $A_{0'} = A_0 +
v A_x$ one gets the action
\begin{equation}S_{EE}  = {k \over 4 \pi} \int \left(  \partial_t  + v \partial_x \right)  \phi \partial_x \phi dx dt \
.
\label{cb}
\end{equation}
Here $v$ is the velocity of the chiral boson. The Hamiltonian
derived from eq. (\ref{cb}) is given by
\begin{equation}H = - {k v \over 4 \pi} \int_0^{2 \pi R} \partial_x \phi \partial_x \phi d x \ ;
\label{hbos}
\end{equation}
the chiral current is  given by $$J = - 1/(2 \pi ) \partial_x \phi
= 1/(2 \pi) \sum_{n = -\infty}^\infty  \alpha_n {\rm exp}
(in(\theta -vt))$$ and the generators of the Virasoro algebra may
be written as
\begin{eqnarray}\left[ \alpha_n, \alpha_m \right] &=& n \delta_{n+m,0} \ ; \nonumber \\
L_n &=& {k \over 2}  \sum_{l = -\infty}^\infty  : \alpha_{n-l} \alpha_l: \quad  ;\nonumber \\
  \left[L_n, L_m \right]  &=& (n-m) L_{n+m} + {c\over 12} (n^3 -n) \delta_{n+m,0} ,
\label{adv}
\end{eqnarray}
with central charge $c =1$.

To get the full $W_{1+\infty}$ algebra one may use the equivalence
between chiral boson and Weyl  fermion in 1+1 dimensions
\cite{jflo}. The chiral boson is, in fact, equivalent \cite{jflo}
to a Weyl fermion $\psi$ described by the Hamiltonian:
\begin{equation}H =  {k v \over 8 \pi} \int_0^{2 \pi R} \psi^\dagger ( i \partial_x ) \psi  d x  + h.c. \ ,
\label{hfer}
\end{equation}
with $\psi (\theta) = 1/\sqrt{R}  \sum_{l = -\infty}^\infty
e^{i(l-1/2)\theta}b_l$. $b_l$ and $b^\dagger_l$ are fermionic
creation and annihilation operator satisfying  $\{b_l, b_m\} =
\delta_{l,m}$. In \cite{ctz} it was shown that it is possible to
construct the  generators of the $W_{1+\infty}$ algebra  in terms
of these fermionic fields:
\begin{eqnarray}V^"_n &=& \int_0^{2\pi} d\theta  \psi^\dagger(\theta) :e^{-in\theta} (i\partial_\theta)^": \psi (\theta) \nonumber  \\
&=&   \sum_{l = -\infty}^\infty  p(l,n,i) b^\dagger_{l-n} b_n \ , i\geq 0 \ ,
\label{gwf}
\end{eqnarray}
where the normal ordering is done in such a way that
$V_n^{i\dagger} = V_{-n}"$ and $p(l,n,i)$ are the ith order
polynomials in $l$. These generators satisfy the algebra
(\ref{gaq}) with $c = 1$ and the generators:  $V_n^0$,  $V_0^1$
form the Virasoro sub-algebra. The new and, at this point rather
simple, remark is that the Hamiltonian (\ref{hfer}) commutes with
{\it all the generators} (\ref{gwf}) and not only with $V_n^0$
and $V_0^1$:
\begin{equation} \left[V^"_n , H \right]  = 0 \ .
\label{paw}
\end{equation}

The CS coefficient $k$ enters the algebra by normalizing the
operators $V_n^0$ and measures the physical charge of the edge
excitations since they satisfy the Kac-Moody algebra
\begin{equation}\left [V^0_n, V^0_m \right] = k n \delta_{n+m,0} \ .
\label{fce}
\end{equation}
For quantum Hall fluids the CS coefficient of the effective theory
is given by $k = 1/\nu$, where $\nu$ is the filling fraction
\cite{Lau} and $k=1$ corresponds to the filled Landau
level. Furthermore, determining the spectrum of the operators
$-V_0^0$ and $V_0^1$ allows to establish that the CS coefficient
$k$ contributes also to the charge Q and the spin J of the bulk
excitations as $Q = q/k$ and $J = q^2/(2k)$.

We want now to analyze the case of the mixed CS theory (BF) relevant for topological superconductivity, where P-and-T symmetry are preserved, on a manifold with boundaries.  Our aim is to compute, via the holographic partition function \cite{ffn},  the topological entanglement entropy since this quantity can distinguish between model described by a {\it fundamental BF} theory from model that are described by an {\it effective BF} theory like the one described in \cite{hansson} after SSB. States of a Chern-Simons theory are accounted for by the conformal blocks of a conformal Þeld theory (CFT). Primary fields of the CFT at the edge are in one-to-one correspondences with the quasiparticles in the topological phase, and  Chern-Simons states may be identiÞed
with the characters of the CFT. Trough the Verlinde formula \cite{verlinde}  one can write the total quantum dimension ${\cal D}$ as
\begin{equation}
{\cal D} = {1 \over S_0^0} \ ,
\label{modent}
\end{equation}
where $ S_0^0$ is is an element of the modular $S$ matrix corresponding to the largest eigenvalue \cite{ffn}.

This matrix has been computed in \cite{stromi} for the mixed CS theory and we have
\begin{equation}
 S_{\beta}^{\beta'} =  {1 \over \sqrt{|\Lambda / \Lambda^*|} }\exp - 2 \pi i (\beta,\beta^`) \ ,
 \label{modulars}
 \end{equation}
 and modular partition function
 \begin{equation}
 Z  = \sum_ {\beta \in \Lambda / \Lambda^*} \chi_\beta \chi_{\bar \beta} \ .
 \label{modularp}
 \end{equation}
 Here $ \chi_\beta$ are the characters and are given by the $(k_1 k_2)^2$ theta functions that solve the quasi-periodic condition that determine the ground state wave functions of the BF model with symmetry $U(1) \times U(1)$ with both topological defects in a dense phase.
 The lattice $\Lambda / \Lambda^*$  is a lattice whose points correspond to the sectors of the rational CFT at the edges (for $k$ rational). $(\beta,\beta^`) $ is a quadratic form \cite{stromi} such that $\exp - 2 \pi i (0,0^`) = 1$.
>From this consideration we see that although formally equivalent, namely two chiral bosons with opposite chirality, the edge theory of topological superconductors and superconductors arising from SSB leads two different topological entanglement entropy. In fact for the former we have, in the superconducting phase only $(k_1 k_2)$ independent theta functions with a topological entanglement entropy $\gamma = \log k$, while the latter has $(k_1 k_2)^2$  independent theta functions and thus a topological entanglement entropy $\gamma = 2\log k$.

The algebra of the edge excitations for PT invariant
incompressible fluids with a gap arising from SSB was analyzed in
\cite{pima}. Due to the PT-invariance, the algebra is still the
direct product of two $W_{1+\infty}$ algebras of opposite
chirality, but  one has to take into account that, as a result
of the fact that the $U(1)$ gauge symmetry is spontaneously
broken, the charge ceases to be a good quantum number in the
ground state of the broken phase. Since the quantum $W_{1+\infty}$
algebra contains a $\hat U(1)$ Kac-Moody current $V_n^0$, one can
identify the electric charge current with the diagonal vector
current $V_n^0 + \bar V_n^0$. This Kac-Moody symmetry has to be
divided by the dynamical symmetry group, yielding the a coset
algebra given by
 \begin{equation} W = {W_{1+\infty} \otimes \bar W_{1+\infty} \over \hat U(1)_{\rm diagonal}} \ ,
\label{ants}
\end{equation}
while for gap of topological origin we will have:
\begin{equation} W = W_{1+\infty} \otimes \bar W_{1+\infty} \ .
\label{ats}
\end{equation}
It should be noticed that, for $c \geq 2$ for both (\ref{ants}) and (\ref{ats}), there are \cite{pima} ground states with the residual
dynamical symmetry
 \begin{equation} W = W_m \otimes \bar W_m \ ,
\label{rcds}
\end{equation}
where $W_m$ is the Fateev, Lykyanov, Zamolodchikov algebra
\cite{flz} in the limit $C_{W_m} \rightarrow m -1$. These ground
states describe unconventional superconductors whose excitation
spectrum consists entirely of neutral, spinon excitations with
non-Abelian fractional statistics and $SU(m)$ isospin symmetry. It
should be observed that the $SU(m)$ symmetry of these excitations
is different from the usual  symmetry of, say, the quark model of
strong interactions \cite{lge} since spinons do not come in the
full $SU(m)$ multiplets but, rather, only the highest-weight state
is present. However, they combine according to the usual $SU(m)$
fusion rules, which explains the non-Abelian character of their
monodromies.

The examples discussed above clarify the fact that charge-spin
separation and non-Abelian statistics are universal properties of
2D superconductors emerging only from incompressibility; the way
in which incompressibility ( the gap) emerges- from topological
order or SSB- affects, in fact, only the algebra of the edge
excitations of the 2D incompressible fluid.
It should by pointed out, however, that  topological superconductors, described by Abelian mixed CS term, correspond to the generic representation of the
$ W_{1+\infty}$ algebra. It has been pointed out in \cite{barwen} that, when  an extra $Z_2$ symmetry is added, namely when the mixed CS term has a $U(1) \times U(1) \times Z_2$ symmetry, the excitations have non-Abelian statistics. Possible applications to topological superconductors are under investigation.

\section{Summary and Concluding Remarks}
 In this paper we showed that, for superconducting systems with non trivial topologies,
incompressibility alone suffices to guarantee both a ground state
degeneracy and fractional statistics.

More important, we showed that the ground state
degeneracy encodes relevant information about the microscopic
mechanism originating the superconducting gap. In fact, when the
gap emerges from SSB, the degeneracy is determined by the symmetry
breaking $U(1) \rightarrow Z_N$ and, thus, the Aharonov-Bohm
interaction between charges and vortices is described by an
effective mixed CS model with degeneracy $(k_1 \times k_2)^2$ if
$k = k_1 / k_2$ is the CS coefficient. At variance, in the
superconduting phase of a topological superconductor described by
the BF model, there is only an electric condensate and the
resulting ground state degeneracy is given by $k_1 k_2$. This
difference in the ground state degeneracy implies that the
entanglement entropy of topological superconductors is half the
one of standard superconductors. A measure of entanglement entropy
may thus distinguish between the two different mechanisms giving
rise to superconductivity.

Finally, since the BF model introduced in \cite{nsm} describes
also superinsulators in terms of a parity and time reversal
invariant topological model, one may wonder if the analysis
carried in this paper may be relevant to characterize relevant
features of topological insulators \cite{mele} also; a topological
insulator being a material which, while being an insulator in the
bulk, allows for the motion of charges at its boundary yielding
edge states which are topologically protected due to time-reversal
invariance.

\end{document}